\documentclass[12pt,a4paper]{article}
\usepackage{graphicx}
\usepackage{amsmath}
\usepackage{amssymb}

\newcommand{\diag}{{\rm diag}}
\newcommand{\Journal}[5]{#1~#2~\textbf{#3}, #5 (#4)}

\begin{document}
\title{Process $\gamma^* \gamma \rightarrow \sigma$ at large virtuality of $\gamma^*$}
\author{
M.~K.~Volkov, A.~E.~Radzhabov, V.~L.~Yudichev
\\[5mm]
\itshape Joint Institute for Nuclear Research, Dubna, Russia \\
}
\date{}
\maketitle
\abstract{The process $\gamma^* \gamma \rightarrow \sigma$ is investigated in the framework of the $SU(2)\times SU(2)$ chiral NJL model. The form factor of the process is derived for arbitrary virtuality of $\gamma^\ast$ in the Euclidean kinematic domain. The asymptotic behaviour of this form factor resembles the asymptotic behaviour of the $\gamma^* \gamma \rightarrow \pi$ form factor.}
\\[5mm]

A recent analysis of the experimental data on the pseudoscalar mesons production in the transition process
$\gamma^\ast\gamma\to P$ \cite{CLEO} (where $\gamma^\ast$, $\gamma$ are photons with the large and small virtuality,
respectively, and $P$ is a pseudoscalar meson) revealed satisfactory agreement with the pertrubative QCD (pQCD)
predictions \cite{Brodsky,Chernyak,MikhRad90,MihRad92,RadRus96} for the asymptotic behaviour of the $\gamma^\ast\gamma\to P$
form factor at large negative virtuality of one of the photons. However, for a correct description of the asymptotic
behaviour of the transition form factor it is not enough to apply only the pQCD technique, and it is necessary to use
the nonpertrubative sector for determining a distribution amplitude (DA) of quarks in a meson.
In the lowest order of pQCD,  the light-cone Operator Product Expansion (OPE) predicts the high $Q^{2}$ behaviour of the form factor as follows \cite{Brodsky,Chernyak}:
\begin{equation}
\left. {F}_P(q_{1}^{2},q_{2}^{2},p^2=0)\right| _{Q^{2}\rightarrow \infty}
=J_P\left( \omega \right) \frac{f_{P}}{Q^{2}}
+O\left(\frac{\alpha_s}{\pi}\right)
+O\left(\frac{1}{Q^4}\right),  \label{AmplAsympt}
\end{equation}
with the asymptotic coefficient given by
\begin{equation}
J_P\left( \omega \right) =\frac{4}{3} \int\limits_{0}^{1} \frac{dx}{1-\omega^2 (
2x-1)^2}\varphi_P^{A}(x),  \label{J}
\end{equation}
where $\varphi_P^{A}(x)$ is the leading-twist meson light-cone DA
normalized by 
\begin{equation}
\int_{0}^{1} dx\,\varphi_P^{A}(x)=1; 
\end{equation}
$\alpha_s$ is the strong coupling constant; $Q^2$ and $\omega$ are, respectively, the total virtuality of the photons and the asymmetry in their distribution
\begin{equation}
Q^{2}=-(q_{1}^{2}+q_{2}^{2})\geqslant 0, \;\;\;\;{\rm and}\;\;\;\; \omega
=(q_{1}^{2}-q_{2}^{2})/(q_{1}^{2}+q_{2}^{2}),\quad |\omega|\leqslant 1,  \label{Omega}
\end{equation}
where $q_1$,$q_2$ are the momenta of photons; $f_P$ is the meson weak decay constant (for the pion $f_\pi=93$ MeV).

Unfortunately, the determination of $\varphi_P^{A}(x)$ is a rather nontrivial problem and can not be performed in the framework of only pQCD. At asymptotically high $Q^{2}$, DA is $\varphi_P^{A,\rm asympt}(x)$ $=6x(1-x)$ and $J_P^{\rm asympt}\left( \left| \omega \right|=1\right) =2$. The fit of the CLEO data\cite{CLEO} for the pion corresponds to $J_\pi^{\rm CLEO}\left( \left| \omega \right| \approx 1\right) = 1.6\pm 0.3$, indicating that already at moderately high
momenta ($Q^2=8$ GeV$^2$) this value is not too far from its asymptotic limit.

In our previous work \cite{PION}, it has been shown that the calculation  of a process $\gamma^\ast\gamma\to P$ amplitude in the framework of a chiral quark model of the NJL type \cite{ECHA} is free from difficulties connected with the necessity of determining DA.
Our results for the form factor asymptotics agreed with experiment \cite{CLEO}, and with the predictions made in QCD sum rules \cite{MikhRad90,MihRad92,RadRus96} and the instanton induced quark model (IQM) \cite{DOAnTo00}. Also, both the NJL and IQM model allow one to derive the meson DA.

Here, we continue the investigation started in \cite{PION} and consider the scalar isoscalar meson production through the process $\gamma^\ast \gamma \to \sigma$ where $\sigma$ is associated with the lightest scalar isoscalar state $f_0(400-1200)$ \cite{PDG}, and the photon $\gamma^\ast$ is off-shell. Direct observation of $\sigma$ is hardly possible; however, in the low-mass region
it can show itself as a resonance in the pion pair production \cite{ppgg}.

In our paper, all the calculations are performed in the framework of the $SU(2)\times SU(2)$ NJL model. We derive a
gauge-invariant expression for the amplitude of $\gamma^\ast \gamma \to \sigma$ and determine the form factor of the process for which the asymptotic behaviour is studied at large virtualities of $\gamma^\ast$. Notice that the asymptotic behaviour of the $\gamma^\ast \gamma \to \sigma$ and $\gamma^\ast \gamma \to \pi$ form factors is similar.

The interaction of mesons with $u$ and $d$ quarks is described by the following quark-meson Lagrangian:
\begin{eqnarray}
L(\bar{q},q,A,\sigma,\pi)= \bar{q}(x) (i\hat{\partial}-M-eQ\hat{A}(x)+g (\sigma(x)+i\gamma_5\boldsymbol \tau
\boldsymbol \pi(x)) )q(x),
\end{eqnarray}
where $\bar{q}$ and $q$ are the $u$ and $d$ quark fields
\begin{equation}
q(x)=\left(\begin{array}{c} u(x) \\ d(x)\end{array}\right);
\end{equation}
$M$ is the constituent mass matrix, $M=\diag(m_u,m_d)$ ($m_u \approx m_d \approx m$); $\hat{A}=A_\mu \gamma^\mu$ is the
photon field; $\sigma$ stands for the scalar isoscalar meson; $\boldsymbol \pi$ is the pion triplet,$\boldsymbol \tau$
are the Pauli matrices ; $e$ is e.m. charge $e^2/4\pi=1/137$; $Q=1/6(1+3\tau_3)$ is the charge operator; and $g$ is the
constant of a quark-meson interaction \footnote{Following \cite{PION}, we do not take into account $\pi - a_1$
transitions. Therefore, $g_\pi=g_\sigma=g$. } defined by the relation $g^{-2}=4 I_2$ where
\begin{eqnarray}\label{I2}
I_2= \frac{3}{(2\pi)^4}\int \frac{d_E^4 k\; \theta(\Lambda^2-k^2)}{(k^2+m^2)^2}.
\end{eqnarray}
The subscript $E$ in $d^4_E k$ means that the integration is performed for momenta in the Euclidean
metric.  The cut-off $\Lambda$ eliminates the UV divergence.

Using the Goldberger-Treiman relation $g=m/f_\pi$
and the relation $g_\rho =
\sqrt{6} g$ \cite{ECHA}, where $g_\rho \approx 6.1$ is the constant describing the decay $\rho\to 2 \pi$, we determine the constituent quark mass $m=234$ MeV and $\Lambda=1050$ MeV.

Let us note that the  $\gamma^\ast\gamma\to\sigma$ amplitude also describes a two photon decay of $\sigma$. The
amplitudes for radiative decays of scalar mesons ($S$) were obtained in \cite{TmF}.
The amplitude of the process $S\rightarrow \gamma \gamma$ has the form
\begin{eqnarray}
A=e^2T^{\mu\nu}_{S \rightarrow \gamma\gamma}\epsilon_\mu(q_1)\epsilon_\nu(q_2),
\end{eqnarray}
where $\epsilon$ are polarization vectors; $q_1$ and $q_2$ are photon momenta.

\begin{figure}
\begin{center}\includegraphics[scale=0.7]{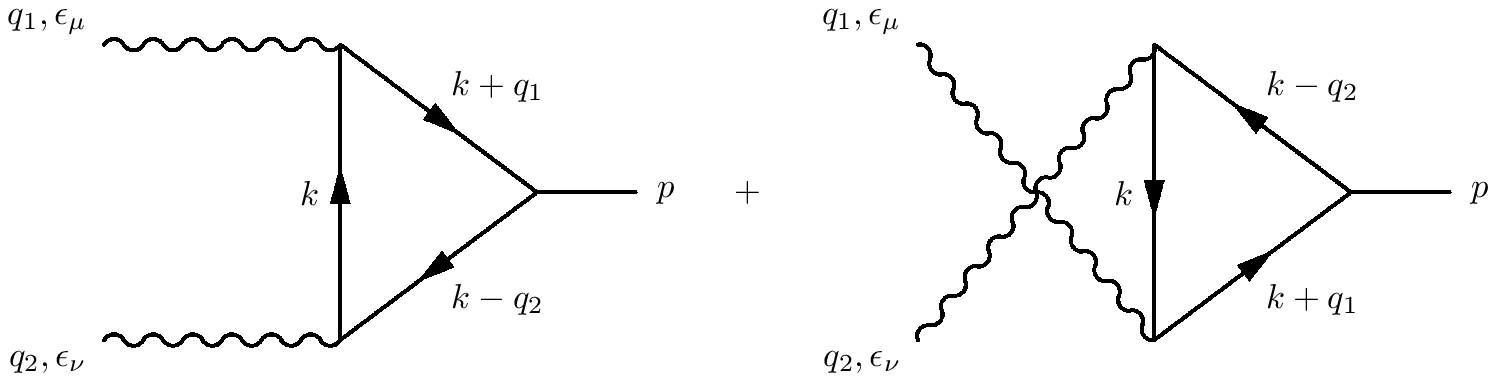}\end{center}
\caption{
Diagrams contributing to the amplitude of the process $\gamma^* \gamma\to \sigma$}
\label{diag1}
\end{figure}

Up to the one-loop, the process is described by the diagrams in Fig.~\ref{diag1}
that give for the tensor $T^{\mu\nu}_{S \to \gamma\gamma}$
\begin{eqnarray}
T^{\mu\nu}_{S \rightarrow \gamma\gamma} &=& T^{(1)\mu\nu}_{S \rightarrow
\gamma\gamma} +T^{(2)\mu\nu}_{S \rightarrow \gamma\gamma},
\end{eqnarray}
where
\begin{eqnarray}
T^{(1)\mu\nu}_{S \rightarrow \gamma\gamma}&=& C_S \left(q_1^\nu q_2^\mu-g^{\mu\nu}(q_1\cdot q_2)\right) J_m(q_1,q_2), \\
T^{(2)\mu\nu}_{S \rightarrow \gamma\gamma} &=& C_S B^{\mu\nu}(q_1,q_2),
\end{eqnarray}
and the transversality condition $\epsilon_\mu(q_1) q_1^\mu=\epsilon_\nu (q_2)q_2^\nu=0$
is taken into account.
Here
\begin{eqnarray}
J_m(q_1,q_2) &=& - \frac{i}{(2\pi)^4} \int d^4 k \frac{1}{D(0)D(q_1)D(-q_2)}, \label{Jm}\\
B^{\mu\nu}(q_1,q_2) &=&  - \frac{i}{(2\pi)^4} \int d^4 k \frac{g^{\mu\nu}(m^2-k^2)+4k^\mu k^\nu}{D(0)D(q_1)D(-q_2)}, \label{Bmn}\\
D(q)&=&m^2-(k+q)^2.
\end{eqnarray}
The constant $C_S$ is to be specified for each type of scalar mesons.
In the case of $\sigma$, it is
\begin{equation}C_\sigma=\frac{40}{3}g m.\end{equation}

The term $T^{(1)\mu\nu}_{\sigma \to \gamma\gamma}$ has a gauge-invariant form.
The Pauli-Villars regularization is necessary
to restore the gauge invariance of $T^{(2)\mu\nu}_{\sigma \to \gamma\gamma}$ which leads to the following form of
$B^{\mu\nu}$ (see Appendix for details) for $q_1^2=q_2^2=0$:
\begin{eqnarray}
B^{\mu\nu}(q_1,q_2)&=& 2 \left(\frac{q_1^\nu q_2^\mu}{(q_1\cdot
q_2)}-g^{\mu\nu}\right)m^2(J_m(0,0)-J_m(q_1,q_2)).
\end{eqnarray}

In our paper, we apply the Pauli-Villars regularization with one subtraction for an arbitrary function $Y(m^2)$ as follows
\begin{eqnarray}
Y^{\mathrm PV}(m^2)= Y(m^2) - Y(M^2)
\end{eqnarray}
Formally, as all integrals we calculate here converge, the regularization can be lifted off by reaching the $M^2\rightarrow \infty$ limit. In our particular case, the regularized expression will, however, differ from the non-regularized one, the constant term, which violates gauge invariance, is thereby cancelled.

In the case of $\sigma$ production through
$\gamma^* \gamma \rightarrow \sigma$, the amplitude can also be divided in two parts
\begin{eqnarray}
T^{\mu\nu}_{\gamma^* \gamma \rightarrow \sigma} &=& T^{(1)\mu\nu}_{\gamma^*
\gamma \rightarrow \sigma} +T^{(2)\mu\nu}_{\gamma^* \gamma \rightarrow \sigma},
\end{eqnarray}
where the first term is again gauge-invariant
\begin{eqnarray}
T^{(1)\mu\nu}_{\gamma^* \gamma \rightarrow \sigma}&=& C_\sigma \left(q_1^\nu
q_2^\mu-g^{\mu\nu}(q_1\cdot q_2)\right) J_m(q_1,q_2),
\end{eqnarray}
whereas the second one becomes gauge invariant after the Pauli-Villars regularization is implemented and then lifted off.
Moreover, as one of the photons is off-shell, $B^{\mu\nu}$ acquires an  additional term
proportional to $q_1^2$ (see Appendix):
\begin{eqnarray}
T^{(2)\mu\nu}_{\gamma^* \gamma \rightarrow \sigma} &=&
 C_\sigma \left(\frac{q_1^\nu q_2^\mu}{(q_1\cdot q_2)}-g^{\mu\nu}\right)
 \Biggl[2m^2\left(J_m(0,0)-J_m(q_1,q_2)\right)-\nonumber\\
&&\left.-\frac{q_1^2}{2 (q_1\cdot q_2)}\frac{i}{(2\pi)^4}\int d^4 k \frac{1}
{D(q_1)}\left( \frac{1} {D(-q_2)}- \frac{1}{D(0)}\right)\right].
\end{eqnarray}
After the gauge invariance of the amplitude is restored, we define the process form factor $F(q_1,q_2)$:
\begin{eqnarray}
T^{\mu\nu}_{\gamma^* \gamma \rightarrow \sigma} &=& (q_1^\nu q_2^\mu- (q_1\cdot
q_2)g^{\mu\nu})F(q_1,q_2) ,
\end{eqnarray}
where
\begin{eqnarray}
&&F(q_1,q_2)=\frac{40 f_\pi g^2}{3}
\left[J_m(q_1,q_2)-\frac{2m^2}{(q_1 \cdot q_2)} (J_m(q_1,q_2)-J_m(0,0))-\right.\nonumber\\
&&\quad \quad \left.- \frac{q_1^2}{2(q_1 \cdot q_2)^2}\frac{i}{(2\pi)^4}\int d^4 k \frac{1}
{D(q_1)}\left(\frac{1}{D(-q_2)}-\frac{1} {D(0)}\right)\right].
\end{eqnarray}
After the change of variable $k = k^\prime-(q_1-q_2)/2$ (in the following we omit the prime symbol), the form factor takes the form
\begin{eqnarray}
&&F(q_1,q_2)= \frac{5 f_\pi g^2}{6\pi^2} \int \frac{d^4 k}{i \pi^2} 
\left[\frac{1-2 m^2/(q_1 \cdot
q_2)}
{D(-p/2)D(p/2)D(-(q_1-q_2)/2)}\right.\nonumber\\
&&+
\frac{2 m^2}{(q_1 \cdot q_2)} \frac{1} {D(0)^3}
\quad\quad+\left. \frac{q_1^2}{2(q_1 \cdot q_2)^2} \frac{1} {D(p/2)} \left(\frac{1} {D(-p/2)}- \frac{1}
{D(-(q_1-q_2)/2)}\right)\right],\nonumber
\end{eqnarray}
where $p=q_1+q_2$. It is convenient to consider $F(q_1,q_2)$ in terms of 
$p^2$ (notice that $p^2=M_\sigma^2$) , $q^2_1$, and $q_2^2$. Further, we restrict ourselves to
the case of $q_2^2=0$; thus, we have two independent variables $p^2$ and $q_1^2$.
The kinematics of the process under investigation
corresponds to large negative $q_1^2$, and we introduce,
for convenience, a positive magnitude $Q^2$ defined as $Q^2=-q_1^2$.
After that, the form factor $F(q_1,q_2)$ can be considered as depending on $p^2$ and $Q^2$
\begin{equation}
\tilde{F}(p^2,Q^2) = F(q_1, q_2)|_{q_2^2=0;q_1^2<0},
\end{equation}

\begin{eqnarray}\label{Fpq}
\tilde{F}(p^2,Q^2)&=& \frac{5 f_\pi g^2}{6\pi^2} \left[\left(1-\frac{4 m^2}{p^2+Q^2}\right)K_\Lambda(p^2,Q^2)\right.+\frac{4
m^2}{p^2+Q^2} L_\Lambda -
\nonumber\\
&&
\left.
-\frac{2 Q^2}{(p^2+Q^2)^2}\left(M_\Lambda(p^2)- N_\Lambda(p^2,Q^2)\right)\right],
\end{eqnarray}
where
\begin{eqnarray}
K_\Lambda(p^2,Q^2) &=& \int^\Lambda \frac{d^4 k}{i \pi^2} \frac{1} {D(-p/2)D(p/2)D(-(q_1-q_2)/2)},  \\
L_\Lambda &=& \int^\Lambda \frac{d^4 k}{i \pi^2} \frac{1} {D(0)^3},  \\
M_\Lambda(p^2)&=& \int^\Lambda \frac{d^4 k}{i \pi^2} \frac{1} {D(-p/2)D(p/2)}, \\
N_\Lambda(p^2,Q^2)&=& \int^\Lambda \frac{d^4 k}{i \pi^2} \frac{1} {D(-(q_1-q_2)/2)D(p/2)}.
\end{eqnarray}
In all integrals, we implement the UV cut-off $\Lambda$ that constrains quark momenta to
the domain where chiral symmetry is spontaneously broken and
the bosonization of quarks takes place(see (\ref{I2})).

An analogous method has been used in \cite{PION}. One can compare (\ref{Fpq}) with the pion form factor obtained in
\cite{PION} for $q_1^2 \neq 0$, $q_2^2 \neq 0$
\begin{eqnarray}
F_\pi(p^2,q_1^2,q_2^2)=C_\pi \int \frac{d ^4 k}{i\pi^2} \frac{1}
{D(-p/2)D(p/2)D(-(q_1-q_2)/2)},
\end{eqnarray}
where $C_\pi$ is a constant.
In \cite{PION},
the following expression was obtained for
$K_\Lambda(p^2,Q^2)$ after introducing Feynman parameters, integrating over the\\ angles, and changing the variables $k^2=u$:
\begin{eqnarray}
&&K_\Lambda(p^2,Q^2)=\int\limits_{0}^{\Lambda^2}\frac{ du\,
u}{m^{2}+u-\frac{p^{2}}{4}}\times\\
&&\times\int\limits_{0}^{1}dx \left[ \frac{1}{\sqrt{b^{2}-a_{+}}\left( b+\sqrt{ b^{2}-a_{+}} \right)
}+\frac{1}{\sqrt{b^{2}-a_{-}}\left( b+\sqrt{ b^{2}-a_{-}} \right) }\right].\nonumber
\end{eqnarray}
Here
\begin{eqnarray}
&&b=m^{2}+u+\frac{1}{2}x Q^{2}-\frac{1}{4}\left( 1-2x \right) p^{2},\nonumber\\
&& a_{\pm }=2ux Q^{2}\left( x \pm \left( 1-x\right) \right) -\left( 1-2x \right) up^{2}. \label{notas2}
\end{eqnarray}
Analogously, one can write the integrals $N_\Lambda(p^2,Q^2)$,$M_\Lambda(p^2)$ as follows:
\begin{eqnarray}
N_\Lambda(p^2,Q^2)&=&2\int\limits_{0}^{\Lambda^2}du \, u \int\limits_{0}^{1}dx
\frac{1}{\sqrt{b^{2}-a_{+}} \left( b+\sqrt{ b^{2}-a_{+}} \right) }, \\
M_\Lambda(p^2)&=&2\int\limits_{0}^{\Lambda^2}
\frac{du \, u}{b_0\left( b_0+\sqrt{ b_0^{2}-a_{+0}} \right) }.
\end{eqnarray}
Here
\begin{eqnarray}
&&b_0=b |_{x \rightarrow 0}=m^{2}+u-\frac{1}{4}p^{2},\quad a_{+0}=a_{+} |_{x \rightarrow 0}=-up^{2}.
\end{eqnarray}

Now we can calculate the asymptotics for $\tilde{F}(p^2,Q^2)$ when $Q^2\to \infty$. Using the approximation described in
\cite{PION} and expanding in a series of $1/Q^2$, we obtain for $K_\Lambda(p^2,Q^2)$, $L_\Lambda$ ,
$M_\Lambda(p^2)$ and $N_\Lambda(p^2,Q^2)$ at $p^2=0$
\begin{eqnarray}
&&K_\Lambda(p^2=0,Q^2)=\frac{1}{Q^2}\int\limits_{0}^{\Lambda^2/m^2 } \frac{du} {1+u}
\ln \left( 1+2u\right) + O\left(\frac{1}{Q^4}\right),
\\
&&L_\Lambda=\frac{1}{m^2}\int\limits_{0}^{\Lambda^2/m^2 } \frac{du\,u} {(1+u)^3},\\
&&M_\Lambda(p^2=0)=\int\limits_{0}^{\Lambda^2/m^2 } \frac{du\,u} {(1+u)^2},\\
&&N_\Lambda(p^2=0,Q^2)=\frac{2 m^2}{Q^2}\int\limits_{0}^{\Lambda^2/m^2 }du\,\ln \left( 1+u\right)  +
O\left(\frac{1}{Q^4}\right).
\end{eqnarray}
Note that the last term is of the order of $1/Q^4$. As a result, the first three terms give us the following asymptotics for the form factor $\tilde{F}(p^2=0,Q^2)$:
\begin{eqnarray}
\tilde{F}(p^2=0,Q^2)&=&J_S\frac{f_\pi}{Q^2} + O\left(\frac{1}{Q^4}\right), \\
J_S &\approx& 3.28+0.96-2.22=2.02.
\end{eqnarray}

\begin{figure}
\begin{center}\includegraphics[scale=0.8]{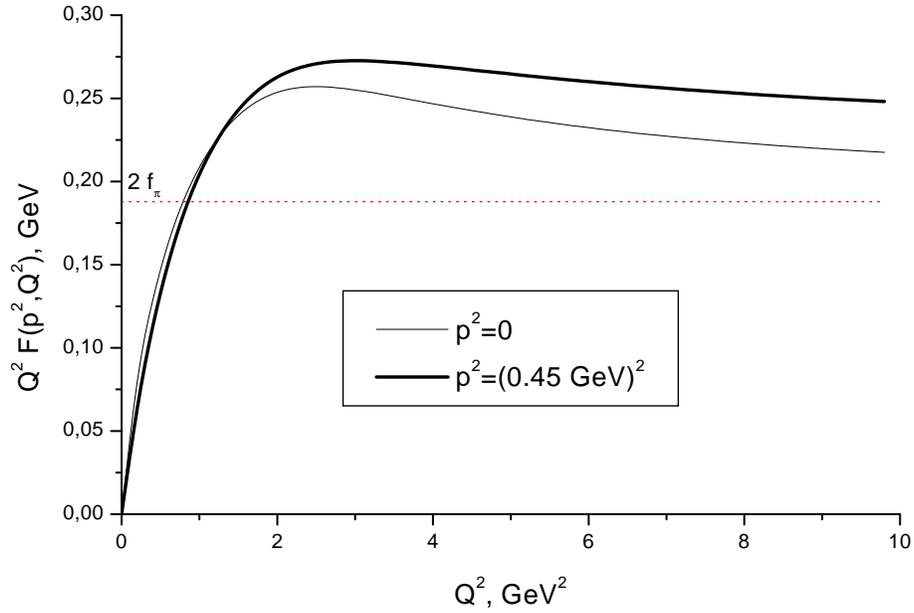}\end{center}
\caption{
The form factor of the process $\gamma^* \gamma\to \sigma$ multiplied by $Q^2$ at $Q^2$ from 0 to 10 GeV$^2$. The thick curve is for $p^2=(0.45$ GeV$)^2$ and the thin is for $p^2=0$. We compare these curves with the theoretical limit for the pion transition form factor: $2f_\pi$ }
\label{ff}
\end{figure}

We also calculate numerically $Q^2 \tilde{F}(p^2,Q^2)$  for intermediate values of $Q^2$
and for a different choice of $p^2$. The curves drawn for $p^2=0$, $p^2=M_\sigma^2=(0.45$ GeV$)^2$ are shown in Fig.~\ref{ff}. The value $M_\sigma=450$ MeV is consistent with recent experimental \cite{Aitala} and theoretical \cite{glue} data. A comparison of the obtained results for the function $Q^2 \tilde{F}(p^2,Q^2)$ with an analogous function for the pion \cite{PION} shows similarity in their asymptotic behaviour. However, some differences take place at intermediate values of $Q^2$. Indeed, the pion function grows monotonically in the whole region of $Q^2$. At the beginning it grows rapidly and after $Q^2>2$GeV$^2$ it slowly approaches the asymptotic value.
In the case of the $\sigma$ meson the function $Q^2 \tilde{F}(p^2,Q^2)$ also experiences a rapidly growth up to approximately $2$GeV$^2$; after that it is slowly decreasing to an asymptotic value. Asymptotic values for these functions are practically equal to each other.

In this work, we have shown that in the framework of the $SU(2)\times SU(2)$ chiral quark model of the NJL type, it is possible to describe the behaviour of the $\gamma^\ast\gamma\to\sigma$ form factor in a wide region of $Q^2$.
The exact expression for the form factor of the process is obtained, and its asymptotic behaviour is investigated.
A comparison of the $\pi$ (see \cite{PION}) and $\sigma$ form factors
reveals similarity in their asymptotic behaviour, which may be understood as a consequence of chiral symmetry.

Our results can be useful in investigations of the processes where  pion pairs in the $S$-wave are produced in two photon collisions \cite{Teryaev}. Indeed, in these processes, as a rule, the $\sigma$-pole diagram plays the dominant role \cite{ECHA,ppgg}.
Some data on the pair pion production is already available \cite{CLEO2}. Also, one can find a discussion on this topic in \cite{Vogt02}.

Notice also that the information concerning the amplitudes $\gamma^\ast \gamma \rightarrow \pi$,$\gamma^\ast \gamma \rightarrow \sigma$ can allow us to calculate corrections to the muon anomalous magnetic moment from the processes $\gamma^\ast \gamma \rightarrow \pi  \rightarrow \gamma^\ast \gamma^\ast$,$\gamma^\ast \gamma \rightarrow \sigma \rightarrow \gamma^\ast\gamma^\ast$, where $\gamma^\ast$ interact with the muon. Last year, this topic has been discussed in various papers (see \cite{ggpsgg}).

Further, we plan to calculate the transition form factor  for
two off-shell photons with arbitrary virtualities in the framework of
both the NJL and IQM model, and it allow us also to define DA.

The authors thank A.E. Dorokhov, S.B. Gerasimov and E.A. Kuraev for fruitful discussions.
The work has been supported by RFBR Grant 02-02-16194.
\clearpage
\appendix

\section*{Appendix}

Let us rewrite the expression for $J_m(q_1,q_2)$ (\ref{Jm}), using the Feynman parameterization
\begin{eqnarray}
J_m(q_1,q_2)&=&- \frac{i}{(2\pi)^4} \int d^4 k \frac{1}
{D(0)D(q_1)D(-q_2)}=\\
&=&- \frac{2i}{(2\pi)^4} \int\limits^1_0 dx \int\limits^{1-x}_0 dy \int \frac{d^4 k}
{(m^2-k^2-2k(q_1x-q_2y)-q_1^2x)^3}.\nonumber
\end{eqnarray}
Shifting the variable $k \rightarrow k - (q_1x-q_2 y) $ and
integrating over momenta and $y$, we obtain
\begin{eqnarray}
J_m(q_1,q_2)=- \frac{1}{(4\pi)^2} \frac{1}{2(q_1 \cdot q_2)}\int\limits_0^1
\frac{dx}{x} \ln \left(\frac{m^2-p^2 x(1-x)}{m^2-q_1^2 x(1-x)}\right).
\label{JmnK}
\end{eqnarray}
Let us rewrite the expression for $B^{\mu\nu}$ (see (\ref{Bmn})), using the Feynman parameters
\begin{eqnarray}
&&B^{\mu\nu}(q_1,q_2) =  - \frac{i}{(2\pi)^4}
 \int d^4 k \frac{g^{\mu\nu}(m^2-k^2)+4k^\mu k^\nu}{D(0)D(q_1)D(-q_2)} =\\
&&\quad= - \frac{2i}{(2\pi)^4}\int\limits^1_0 dx \int\limits^{1-x}_0 dy \int d^4 k \frac{g^{\mu\nu}(m^2-k^2)+4k^\mu
k^\nu} {(m^2-k^2-2k(q_1x-q_2y)-q_1^2x)^3}.\nonumber
\end{eqnarray}
Again,  after the change of variables, we have for $B^{\mu\nu}$
\begin{eqnarray}
&&B^{\mu\nu}(q_1,q_2) =- \frac{2 i}{(2\pi)^4} \int\limits^1_0 dx \int\limits^{1-x}_0 dy \int d^4 k 
\frac{A^{\mu\nu}} {(m^2-k^2-q_1^2x(1-x)-2(q_1\cdot q_2)x y)^3},  \nonumber\\ 
&&A^{\mu\nu}=g^{\mu\nu}(m^2-k^2+2k\cdot (q_1x-q_2 y)-q_1^2x^2+2(q_1\cdot q_2)x y)+ \nonumber\\
&&\quad\quad\quad+4(k^\mu k^\nu-k ^\mu (q_1x-q_2 y)^\nu-(q_1x-q_2 y)^\mu k^\nu+\\ 
&&
\quad\quad\quad+(q_1x-q_2 y)^\mu (q_1x-q_2 y)^\nu). \nonumber
\end{eqnarray}
The integration over momenta gives
\begin{eqnarray}\label{Bmunu}
&&B^{\mu\nu}(q_1,q_2) =\frac{g^{\mu\nu}}{(4\pi)^2} \int\limits^1_0 dx \int\limits^{1-x}_0 dy \frac{m^2-q_1^2 x^2
+ 2(q_1\cdot q_2)x y} {m^2-q_1^2x(1-x)-2(q_1 \cdot q_2)x y}-\nonumber\\ 
&&
\quad\quad\quad -\frac{4 q_1^\nu q_2^\mu }{(4\pi)^2} \int\limits^1_0 dx \int\limits^{1-x}_0 dy \frac{x y} {m^2-q_1^2x(1-x)-2(q_1 \cdot q_2)x y}.
\end{eqnarray}
Let us denote the denominator in (\ref{Bmunu}) as $Z$
\begin{equation}
Z=m^2-q_1^2x(1-x)-2(q_1 \cdot q_2)x y.
\end{equation}
Using this equation, we substitute the mass squared $m^2$ in the numerator of (\ref{Bmunu})
\begin{eqnarray}\label{Bmunu3}
&&B^{\mu\nu}(q_1,q_2) =\frac{g^{\mu\nu}}{(4\pi)^2} \int\limits^1_0 dx \int\limits^{1-x}_0 dy \frac{Z + 4(q_1\cdot
q_2)x y +q_1^2x(1-2x)} {Z}-\nonumber\\
&&\quad \quad\quad\quad\quad-\frac{q_1^\nu q_2^\mu}{(4\pi)^2} \int\limits^1_0 dx \int\limits^{1-x}_0 dy \frac{4 x y}
{Z}.
\end{eqnarray}
Now it is easy to separate the gauge-invariant and
noninvariant parts in (\ref{Bmunu3})
\begin{eqnarray}\label{Bmunu2}
B^{\mu\nu}(q_1,q_2) &=& \frac{g^{\mu\nu}(q_1\cdot q_2)-q_1^\nu q_2^\mu}{(4\pi)^2} \int\limits^1_0 dx
\int\limits^{1-x}_0 dy \frac{4 x y} {Z} \\
&&+\frac{g^{\mu\nu}}{(4\pi)^2} \int\limits^1_0 dx
\int\limits^{1-x}_0 dy \frac{ Z+q_1^2x(1-2x)} {Z}.\nonumber
\end{eqnarray}
The first term in (\ref{Bmunu2}) is gauge-invariant, while the other
is noninvariant and can also be divided in two terms
\begin{eqnarray}\label{Bgenv}
\frac{1}{(4\pi)^2} \int\limits^1_0 dx \int\limits^{1-x}_0 dy + \frac{q_1^2}{(4\pi)^2}
\int\limits^1_0 dx \int\limits^{1-x}_0 dy \frac{x(1-2x)} {Z}.
\end{eqnarray}
The first term in (\ref{Bgenv}) is a constant and is cancelled
by the Pauli-Villars regularization.
Integrating over $y$ in the second term, we obtain
\begin{eqnarray}
-\frac{1}{(4\pi)^2} \frac{q_1^2}{2(q_1 \cdot q_2)}\int\limits^1_0 dx (1-2x) \ln
\left(\frac{m^2-(q_1^2+2(q_1 \cdot q_2))x(1-x)}{m^2-q_1^2 x(1-x)}\right).
\end{eqnarray}
After the substitution $1-2x=z$, one can see that this part equals zero
\begin{eqnarray} -\frac{1}{(4\pi)^2} \frac{q_1^2}{4(q_1 \cdot q_2)} \int\limits^1_{-1}
dz z \ln \left(\frac{4 m^2+(q_1^2+2(q_1 \cdot q_2)(1-z^2)}{4
m^2+q_1^2(1-z^2)}\right) = 0.
\end{eqnarray}
Therefore, the expression for $B^{\mu\nu}$ becomes gauge invariant
after the Pauli-Villars regularization is implemented and then lifted off
\begin{eqnarray}
B^{\mu\nu}(q_1,q_2)&=&(q_1^\nu q_2^\mu- (q_1 \cdot q_2)g^{\mu\nu})
(I(m^2)-I(M^2))|_{M^2 \to \infty}, \\
I(m^2)&=&-\frac{1}{(4\pi)^2} \int\limits^1_0 dx \int\limits^{1-x}_0 dy \frac{4 x y}
{Z}.
\end{eqnarray}
Integrating over $y$ in $I(m^2)$ we obtain
\begin{eqnarray}\label{intVs1}
&&I(m^2)=\frac{1}{(4\pi)^2} \frac{1}{(q_1 \cdot q_2)}
+\frac{1}{(4\pi)^2} \frac{m^2}{(q_1 \cdot q_2)^2}\int\limits_0^1  \frac{dx}{x} \ln \left(\frac{m^2-p^2x(1-x)}{m^2-q_1^2 x(1-x)}\right)- \nonumber\\
&&\quad\quad-\frac{1}{(4\pi)^2} \frac{q_1^2}{(q_1 \cdot q_2)^2}\int\limits_0^1 dx (1-x) \ln
\left(\frac{m^2-p^2x(1-x)}{m^2-q_1^2x(1-x)}\right).
\end{eqnarray}
Now, let us show that $I(m^2)$ can be expressed through the integral
\begin{eqnarray}\label{int1}
&&\frac{i}{(2\pi)^4} \int d^4 k \frac{D(0)-D(-q_2)}
{D(0)D(q_1)D(-q_2)}= \nonumber\\
&& = \frac{2i}{(2\pi)^4} \int\limits^1_0 dx \int\limits^{1-x}_0 dy \int d^4 k \frac{(k\cdot q_2)}
{(m^2-k^2-2k(q_1x-q_2y)-q_1^2x)^3}=\nonumber\\
&&=\frac{1}{(4\pi)^2} \int\limits_0^1 dx \ln \left(\frac{m^2-p^2 x(1-x)}{m^2-q_1^2 x(1-x)}\right).
\end{eqnarray}
Noting that the following identity is satisfied:
\begin{eqnarray}\label{int2}
\frac{1}{(4\pi)^2}\int\limits_0^1 dx (1-x)\ln \left(\frac{m^2-p^2 x(1-x)}{m^2-q_1^2
x(1-x)}\right)=
\frac{1}{2(4\pi)^2}\int\limits_0^1 dx \ln \left(\frac{m^2-p^2 x(1-x)}{m^2-q_1^2
x(1-x)}\right),
\end{eqnarray}
we obtain for (\ref{int1})
\begin{eqnarray}
\frac{i}{(2\pi)^4} \int d^4 k \frac{D(0)-D(-q_2)}
{D(0)D(q_1)D(-q_2)}=
\frac{2}{(4\pi)^2}\int\limits_0^1 dx (1-x)\ln \left(\frac{m^2-p^2
x(1-x)}{m^2-q_1^2 x(1-x)}\right).\label{intVs1K}
\end{eqnarray}
Using (\ref{JmnK}) and (\ref{intVs1K}), we rewrite $I(m^2)$ as follows:
\begin{eqnarray}
I(m^2)&=&\frac{1}{(4\pi)^2} \frac{1}{(q_1 \cdot q_2)}-\frac{2m^2}{(q_1 \cdot
q_2)} J_m(q_1,q_2)-\nonumber\\
&&-\frac{q_1^2}{2(q_1 \cdot q_2)^2}\frac{i}{(2\pi)^4} \int d^4
k \frac{D(0)-D(-q_2)} {D(0)D(q_1)D(-q_2)}. \label{Im2}
\end{eqnarray}
Formally, the first term in (\ref{Im2}) can be rewritten as
\begin{eqnarray}
\frac{2m^2}{(q_1 \cdot q_2)}J_m(0,0)
\end{eqnarray}
because
\begin{eqnarray}
J_m(0,0)=-\frac{i}{(2\pi)^4} \int \frac{d^4 k}
{D(0)^3}=\frac{1}{2(4\pi)^2m^2}.
\end{eqnarray}

Finally, the regularized expression for $B^{\mu\nu}(q_1,q_2)$ has the form
\begin{eqnarray}
B^{\mu\nu}(q_1,q_2)&=&(q_1^\nu q_2^\mu- (q_1 \cdot q_2)g^{\mu\nu})
\left(-\frac{2m^2}{(q_1 \cdot q_2)} J_m(q_1,q_2)+\frac{2m^2}{(q_1 \cdot q_2)}J_m(0,0)-\right.\nonumber\\
&&-\left.\frac{q_1^2}{2(q_1 \cdot q_2)^2}\frac{i}{(2\pi)^4}\int d^4 k
\frac{D(0)-D(-q_2)} {D(0)D(q_1)D(-q_2)}\right). \label{BmunuReg}
\end{eqnarray}
Equation (\ref{BmunuReg}) is expressed in terms of formal integrals.
This gives us an advantage of further implementing a regularization different from the Pauli-Villars scheme.

\clearpage

\end{document}